\documentclass[
aps,%
12pt,%
final,%
notitlepage,%
oneside,%
onecolumn,%
nobibnotes,%
nofootinbib,
superscriptaddress,%
noshowpacs,%
centertags]%
{revtex4}
\RequirePackage{graphicx}

\def\refitem#1{\relax}

\begin{document}
\title{Physical mechanism of the (tri)critical point generation}

\author{\firstname{K.A.} \surname{Bugaev}}
\email{Bugaev@th.physik.uni-frankfurt.de}
\affiliation{Bogolyubov Institute for Theoretical Physics, Kiev, 03680, Ukraine}

\author{\firstname{A.I.} \surname{Ivanitskii}}
\email{A_Iv_@ukr.net}
\affiliation{Bogolyubov Institute for Theoretical Physics, Kiev, 03680, Ukraine}

\author{\firstname{E.G.} \surname{Nikonov}}
\email{E.Nikonov@jirn.dubna}
\affiliation{Laboratory for Information Technologies, JINR Dubna, 
141980 Dubna, Russia}

\author{\firstname{V.K} \surname{Petrov}}
\email{Vkpetrov@yandex.ru}
\affiliation{Bogolyubov Institute for Theoretical Physics, Kiev, 03680, Ukraine}

\author{\firstname{A. S.} \surname{Sorin}}
\email{Sorin@theor.jinr.ru}
\affiliation{Bogoliubov  Laboratory of Theoretical Physics, JINR Dubna, 
141980 Dubna, Russia}

\author{\firstname{G.M.} \surname{Zinovjev}}
\email{Gennady.Zinovjev@cern.ch}
\affiliation{Bogolyubov Institute for Theoretical Physics, Kiev, 03680, Ukraine}

\vspace*{-1.8cm}

\maketitle

{
\begin{center}
\small 
\begin{minipage}[b]{15.6cm}
{\bf Abstract.} We discuss some ideas resulting from a phenomenological relation
recently declared between the tension of string connecting the 
static quark-antiquark pair and surface tension of corresponding 
cylindrical bag. This relation analysis leads to the temperature
of vanishing surface tension coefficient of the QGP bags at zero 
baryonic charge density as $T_{\sigma} = 152.9 \pm 4.5$ MeV. We 
develop the view point that this temperature value is not a fortuitous 
coincidence with the temperature of (partial) chiral symmetry
restoration as seen in the lattice QCD simulations. Besides, we 
argue that $T_{\sigma}$ defines the QCD (tri)critical endpoint
temperature and claim that a negative value of surface tension 
coefficient recently discovered is not a sole result, but should also 
exist in 
ordinary liquids at the supercritical temperatures. 
\end{minipage}
\end{center}
}

\vspace*{0.3cm}

{\bf 1. Introduction.}
The contemporary  paradigm   that at the deconfinement region the quark gluon plasma (QGP) is a strongly interacting plasma  \cite{Shuryak:sQGP} seems to become a commonplace fact in the lattice QCD  \cite{fodorkatz}. 
We, however, would like to point out that recently there appeared two almost revolutionary findings related to this paradigm.  
The first of them is found by the lattice QCD Wuppertal-Budapest (WB) group  
\cite{WupBuD:2009} and it states that at vanishing baryonic densities the (partial) chiral symmetry restoration temperature  $T_{chir} $ is between $146 \pm 2 \pm 3 $ and $152 
\pm 3 \pm 3 $ MeV.  This $T_{chir} $ value  is essentially 
smaller than the cross-over temperature $T_{co} = 170 \pm 4 \pm 3$ MeV \cite{WupBuD:2009}.
Indirectly this finding  supports the possibility of the quarkyonic phase existence 
\cite{QY}. 
The second of them is that at $T_{co}$ and  vanishing baryonic densities the surface tension of quark gluon bags is negative \cite{String:09}. As we argue  the latter signals about new
 physics  which, so far, was not investigated by  the theory of ordinary liquids. 
Also here we discuss two kinds   of  exactly solvable statistical  models, the  quark gluon bags with surface tension models,  which incorporate the existence 
of negative surface tension coefficient and use it to generate  the tricritical  
\cite{QGBSTM,QGBSTMb}
and critical \cite{QGBSTM2,QGBSTM2b}
endpoint at some finite value of baryonic 
chemical potential.


{\bf 2. Relation between string tension and surface tension.}
In order to estimate the surface tension of QGP bags let us consider 
the static quark-antiquark pair connected by the unbreakable color tube of length $L$ and radius $R \ll L$.
In the limit of large $L$  the free energy of the color tube is $F_{str} \rightarrow \sigma_{str} L$. The non-vanishing  string tension coefficient $\sigma_{str}$ signals about the color confinement, while its zero value is the usual measure for the deconfinement. 
Now we  consider the same tube as an elongated cylinder of the same radius and length. 
For its free energy we use the standard  parameterization \cite{String:09}
\begin{eqnarray}\label{EqBugaevI}
&&\hspace*{-0.5cm}F_{cyl} (T, L, R)  = 
- p_v(T) \pi R^2 L + \sigma_{surf} (T) 2 \pi R L +  T \tau \ln\left[
\frac{\pi R^2 L }{V_0} \right] \,,
\end{eqnarray}
where $p_v (T)$ is the bulk pressure inside a bag, $\sigma_{surf} (T)$ is the temperature dependent 
surface tension coefficient,  while the 
last term on the right hand side above  is the Fisher topological term \cite{Fisher:67}  which 
is proportional to the Fisher exponent $\tau = const > 1 $ \cite{QGBSTM, QGBSTMb, QGBSTM2, QGBSTM2b} and $V_0 \approx 1$ fm$^3$ is a 
normalization constant.  
Since we consider the same object then its free energies calculated as the color tube  and as  the cylindrical bag should be equal to each other. Then for large separating distances $L \gg R$ one 
finds the following relation 
\begin{equation}\label{EqBugaevII}
\sigma_{str} (T) = \sigma_{surf} (T)\, 2 \pi R~ - ~p _v (T) \pi R^2 + \frac{T \tau}{L} \ln\left[
\frac{\pi R^2 L }{V_0} \right] \rightarrow \sigma_{surf} (T)\, 2 \pi R~ - ~p _v (T) \pi R^2 \,.
\end{equation}
In doing so, in fact,  we match an ensemble of all string shapes of fixed $L$ to a mean elongated cylinder, 
which according to the original Fisher idea \cite{Fisher:67} and the results   of the 
Hills and Dales Model (HDM) \cite{Bugaev:04b,Bugaev:05a} represents a sum of all surface deformations of such  a  bag.
The last equation allows one to determine the $T$-dependence of bag surface tension as
\begin{equation}\label{EqBugaevIII}
\sigma_{surf} (T) = \frac{\sigma_{str} (T)}{ 2 \pi R} ~ + ~ \frac{1}{2} \, p_v (T) R \,,
\end{equation}
if $R(T)$, $\sigma_{str} (T)$ and $p_v (T)$ are known. This relation opens a principal possibility to determine the bag surface tension  directly from the lattice QCD simulations for any $T$. 
Also it allows us to estimate  the surface tension at $T=0$. Thus, taking the typical value of the bag model pressure which is used in hadronic spectroscopy $p_v (T=0) = - (0.25)^4$ GeV$^4$ and inserting 
into (\ref{EqBugaevIII}) the lattice QCD values  $R=0.5$ fm  and $\sigma_{str} (T=0) = (0.42)^2$ GeV$^2$ 
\cite{StrTension2}, one finds  $\sigma_{surf} (T=0) = (0.2229~ {\rm GeV})^3 + 
0.5\, p_v\, R\approx (0.183~{\rm GeV})^3 \approx 157.4$ MeV fm$^{-2}$ \cite{String:09}.

The found value of the bag surface tension at zero temperature is very important for the phenomenological equations of state  of strongly interacting matter in two respects. 
Firstly, according to HDM the obtained value defines the temperature at which    the bag surface tension coefficient changes the sign \cite{Bugaev:04b,Bugaev:05a,Bugaev:10}
\begin{equation}\label{EqBugaevIV}
 T_\sigma = \sigma_{surf} (T=0)\, V_0^\frac{2}{3} \cdot \lambda^{-1}~ \in ~[148.4;~ 157.4] ~{\rm MeV} \, , 
\end{equation}
where the constant $\lambda = 1$ for the Fisher parameterization of the $T$-dependent  surface tension coefficient \cite{Fisher:67} or 
$\lambda \approx 1.06009$, if we use the parameterization derived within the HDM for surface deformations \cite{Bugaev:04b,Bugaev:05a,Bugaev:10}.  
A straightforward  evaluation of the entropy density of the elongated cylinder made from (\ref{EqBugaevI})   in  \cite{String:09} shows that at the cross-over temperature the surface tension coefficient of bag  should be negative
otherwise its   entropy density is negative. 

The remarkable fact  is that the value of temperature $ T_\sigma$ in (\ref{EqBugaevIV}) just  matches that one of 
(partial) chiral symmetry restoration found by the WB group \cite{WupBuD:2009}, i.e. $ T_\sigma = T_{chir}$. 
In other words, two different physical quantities, i.e. the chiral condensate and surface tension coefficient, which are obtained by entirely independent methods indicate  that  the  properties of  strongly interacting matter   are qualitatively   changed in the same temperature range.  Such a `coincidence' can be understood naturally, if we recall that the relevant degrees of freedom (=constituents), interaction between them together with the properties of their surface  are qualitatively  different in different   phases of matter. 
Thus,   one can expect that  different 
physics is indicated by the sign change  of the surface  tension. This conclusion is   supported by the results of  quark gluon bags with surface tension models  \cite{QGBSTM,QGBSTMb,QGBSTM2,QGBSTM2b}, by  the Fisher droplet model (FDM) \cite{Fisher:67}  and by the simplified statistical multifragmentation model (SMM) \cite{SMM1}. 

Secondly, according to the most successful models of liquid-gas phase transition, i.e. FDM  \cite{Fisher:67} and  SMM \cite{Bondorf:95},  the surface tension coefficient linearly depends on temperature. This conclusion is well supported by HDM and by microscopic models of vapor-liquid interfaces  \cite{JGross:09}.
Therefore,   the temperature $T_\sigma$ in (\ref{EqBugaevIV}), at which the surface tension coefficient vanishes,  is also  the  temperature of the (tri)critical endpoint 
$T_{cep}$ of the liquid-gas phase diagram.  On the basis of these results we conclude that  the value of QCD critical endpoint temperature is 
$T_{cep}= T_\sigma =  152.9 \pm 4.5 $ MeV.  Hopefully, the latter can be verified by the lattice QCD simulations using Eq. (\ref{EqBugaevIII}). 




{\bf 3. The role of negative surface tension coefficient.}
The quark gluon bags with surface tension models with the tricritical 
\cite{QGBSTM,QGBSTMb} and critical  \cite{QGBSTM2,QGBSTM2b} point employ the same physical mechanism of the endpoint generation as  FDM \cite{Fisher:67}
and SMM  \cite{SMM1,Bondorf:95}  which is typical for  simple liquids
\cite{Fisher:67}: at the phase coexistence line the difference of bulk parts of free energy of two phases vanishes due to Gibbs criterion, whereas   at the endpoint, additionally,  the surface part of free energy  of liquid phase disappears.  However, in contrast to  FDM and SMM,  in which the surface tension coefficient   is zero above the  endpoint temperature, the  quark gluon bags with surface tension models from the very beginning employ  negative values of  surface tension coefficient above 
$T_{cep}$. So far, an existence of  negative surface tension coefficient above $T_{cep}$ is the only know physical reason  explaining  why the first order phase transition  degenerates into a cross-over \cite{QGBSTM}. 
Now the question is whether  negative surface tension exists in the usual liquids. 
The experimental data on  negative  surface tension coefficient of usual liquids are, of course, unknown. However, if one takes highly accurate  experimental data in the critical endpoint vicinity, then one finds not only that  the surface tension coefficient approaches zero, but, in contrast to the wide spread belief, its full $T$ derivative  does not vanish and remains finite at  $T_{cep}$: $\frac{d \sigma_{surf}}{d ~T} < 0$ \cite{Scaling:06}. Therefore, just the naive extension of these data to the temperatures above  $T_{cep}$ would lead to  negative values of  surface tension coefficient at the supercritical temperatures. On the other hand, if one, as usually,   believes that $\sigma_{surf} = 0$ for $T >T_{cep} $, then  it is absolutely unclear what  physical process can lead to  simultaneous existence of  the discontinuity of $\frac{d \sigma_{surf}}{d ~T}$ at 
$T_{cep}$ and the  smooth behavior   of the pressure's  first and second derivatives   at the cross-over. Therefore, we conclude that  negative surface tension coefficient   at supercritical temperatures is also necessary for ordinary liquids although up to now this question has not been  investigated. 
The quark gluon bags with surface tension models tell us that  the surface tension coefficient is the natural order parameter  allowing  one to distinguish the {\bf  quark gluon liquid} phase which is represented by a single bag of infinite size with $\sigma_{surf} \ge 0$ from the {\bf  QGP} that is the mixture of bags of all sizes which
due to $\sigma_{surf} < 0$ has the  finite  mean volume. 
Also it is clear that  the line $\sigma_{surf} = 0$ is  the natural border between the 
QGP ($\sigma_{surf} < 0$) and hadron gas ($\sigma_{surf} > 0$) at the cross-over region. 

{\bf 4. Conclusions.} Here we discuss the relation between the tension of the color string connecting the static quark-antiquark pair and the surface tension of the corresponding  cylindrical bag. Such a relation allows us to determine the  temperature of vanishing surface tension coefficient of QGP bags at zero baryonic charge  density as  $T_\sigma  =  152.9 \pm 4.5 $ MeV. We  argue  that just 
this range of temperatures does not randomly matches  the range of the (partial) chiral symmetry restoration temperatures  found by the WB collaboration \cite{WupBuD:2009}.
Using Fisher conjecture  \cite{Fisher:67} and  the exact results found for  the temperature dependence of surface tension coefficient  from   the partition of  surface deformations \cite{Bugaev:04b,Bugaev:05a,Bugaev:10},     we conclude that the same temperature range 
corresponds to the  value of QCD (tri)critical endpoint temperature, i.e. $T_{cep}= T_\sigma =  152.9 \pm 4.5 $ MeV.  Furthermore, we claim that the negative values of  surface 
tension coefficient of QGP bags found recently in \cite{String:09}   are not unique, but also should exist for the supercritical temperatures of usual liquids.

\noindent
{\bf Acknowledgments.} E.G.N. and A.S.S. have been supported in part by the Russian Fund for Basic Research under the grant number 11-02-01538a.

\end{document}